%% file: main.tex
\documentclass[aps,prl,twocolumn,noeprint,showpacs,nofootinbib,superscriptaddress]{revtex4}

\usepackage{graphicx}
\usepackage{color}
\usepackage{natbib}
\usepackage{epsfig}
\usepackage{float}
\usepackage{xcolor}

\usepackage{bibunits}

\usepackage{amsmath,amssymb,amsfonts}

\newif\iffull
\fulltrue

\newif\ifversA
\versAfalse

\newif\ifversB
\versBtrue

\input{journalheader}

\begin{document}

\title{Equation of state constraints from the threshold binary mass for prompt collapse of neutron star mergers}

\author{Andreas Bauswein}
\affiliation{GSI Helmholtzzentrum f\"ur Schwerionenforschung, Planckstra{\ss}e 1, 64291 Darmstadt, Germany}

\author{Sebastian Blacker}
\affiliation{GSI Helmholtzzentrum f\"ur Schwerionenforschung, Planckstra{\ss}e 1, 64291 Darmstadt, Germany}
\affiliation{Institut f\"ur Kernphysik, Technische Universit\"at Darmstadt, 64289 Darmstadt, Germany}

\author{Vimal Vijayan}
\affiliation{GSI Helmholtzzentrum f\"ur Schwerionenforschung, Planckstra{\ss}e 1, 64291 Darmstadt, Germany}

\author{Nikolaos Stergioulas}
\affiliation{Department of Physics, Aristotle University of Thessaloniki, 54124 Thessaloniki, Greece}

\author{Katerina Chatziioannou}
\affiliation{Center for Computational Astrophysics, Flatiron Institute, 162 5th Ave, New York, NY 10010, USA}

\author{James A. Clark}
\affiliation{Center for Relativistic Astrophysics, School of Physics, Georgia Institute of Technology, Atlanta, Georgia 30332, USA}

\author{Niels-Uwe F. Bastian}
\affiliation{Institute of Theoretical Physics, University of Wroc{\l}aw, 50-205 Wroc{\l}aw, Poland}

\author{David B. Blaschke}
\affiliation{Institute of Theoretical Physics, University of Wroc{\l}aw, 50-205 Wroc{\l}aw, Poland}
\affiliation{National Research Nuclear University (MEPhI), 115409 Moscow, Russia}
\affiliation{Bogoliubov Laboratory for Theoretical Physics, Joint Institute for Nuclear Research, 141980 Dubna, Russia}

\author{Mateusz Cierniak}
\affiliation{Institute of Theoretical Physics, University of Wroc{\l}aw, 50-205 Wroc{\l}aw, Poland}

\author{Tobias Fischer}
\affiliation{Institute of Theoretical Physics, University of Wroc{\l}aw, 50-205 Wroc{\l}aw, Poland}

\date{\today}

\begin{abstract}
Using hydrodynamical simulations for a large set of high-density matter equations of state (EoSs) we systematically determine the threshold mass $M_\mathrm{thres}$ for prompt black-hole formation in equal-mass and asymmetric neutron star (NS) mergers. We devise the so far most direct, general and accurate method to determine the unknown maximum mass of nonrotating NSs from merger observations revealing $M_\mathrm{thres}$. 
Considering hybrid EoSs with hadron-quark phase transition, we identify a new, observable signature of quark matter in NS mergers. Furthermore, our findings have direct applications in gravitational wave searches, kilonova interpretations and multi-messenger constraints on NS properties.

\end{abstract}


% Gravitaion waves 04.30.−w
% 04.30.Tv Gravitational-wave astrophysics
% 12.38.−t Quantum chromodynamics
% 12.38.Aw General properties of QCD (dynamics, confinement, etc.)
% 12.38.Mh Quark-gluon plasma
% 26.50.+x Nuclear physics aspects of novae, supernovae, and other explosive environments
% 26.60.+c Nuclear matter aspects of neutron stars
% 95.30.Lz Hydrodynamics
% 95.30.Sf Relativity and gravitation
% 95.85.Sz Gravitational radiation, magnetic fields, and other observations
% 97.60.Jd Neutron stars

\maketitle

\begin{bibunit}%[paper]

{\it Motivation and context:}
With the sensitivity increase of current gravitational-wave (GW) detectors, observations of neutron star (NS) mergers will become routine in the very near future~\cite{Abbott2017,Abbott2020}. Also, the identification of electromagnetic counterparts will succeed frequently as sky localizations from the GW signal improve, more dedicated instruments become operational and observing strategies advance. This includes the radiation from ejecta in the ultraviolet, optical and infrared wavebands, so-called kilonovae~\cite{Metzger2019}, but also gamma ray, X-ray and radio emission from relativistic outflows~\cite{Abbott2017b}.

One of the most basic features of a NS coalescence is the immediate merger product, which can either be a black hole (BH) for high total binary masses or a NS remnant for lower total masses~\cite{Shibata2005,Baiotti2008,Hotokezaka2011,Bauswein2013}. The latter may undergo a delayed collapse to a BH. Generally, the NS remnant's lifetime increases with decreasing total binary mass~\cite{Faber2012,Baiotti2017,Paschalidis2017,Friedman2018,Bauswein2019b,Baiotti2019,Duez2019,Lucca2019,Radice2020}.

Based on the distinction between prompt and delayed BH formation for systems with different total binary mass, one can introduce a threshold binary mass $M_\mathrm{thres}$ for direct collapse, which is measurable. The total binary mass $M_\mathrm{tot}$ can be inferred with good precision from the inspiral GW signal, i.e. the premerger phase~\footnote{In practice, the chirp mass is measured with very high precision and constraints on the binary mass ratio are required to obtain $M_\mathrm{tot}$~\cite{Abbott2017,Abbott2019}.}. The merger outcome can be observationally discerned either by the presence of strong postmerger GW emission from a NS remnant~\cite{Clark2014,Torres-Rivas2019} (absent for direct BH formation) or from the properties of the electromagnetic counterpart, which is expected to be relatively dim for prompt-collapse events because of reduced mass ejection~\cite{Bauswein2013a,Hotokezaka2013,Margalit2019}. Thus, a number of measurements with different $M_\mathrm{tot}$ and information on the merger product yields $M_\mathrm{thres}$. The measurement uncertainty essentially depends on how the detections sample the $M_\mathrm{tot}$ range. It should thus continuously decrease with the number of events which allow a distinction between the possible outcomes.

The threshold binary mass is highly important for the interpretation of NS merger observations~\cite{Metzger2017,Ruiz2017,Margalit2019,Coughlin2019a,Paschalidis2019,Lue2019,Abbott2019,Gill2019,Metzger2019,Foley2020,Coughlin2020,Agathos2020,Chen2020,Krueger2020,Abbott2020,Antier2020,Nathanail2020}. Moreover, $M_\mathrm{thres}$ depends in a specific way on the incompletely known equation of state (EoS) of NS matter~\cite{Bauswein2013}. Therefore, understanding the EoS dependence of the collapse behavior is crucial for current and future constraints on unknown properties of high-density matter and of NSs, such as their maximum mass~\cite{Bauswein2013}, radii~\cite{Bauswein2017,Koeppel2019,Capano2020} and tidal deformabilities~\cite{Radice2018,Radice2019,Bauswein2019a,Bauswein2019b}. The prospect to determine $M_\mathrm{max}$ is very notable, where solid lower limits are currently provided by pulsar measurements~\cite{Antoniadis2013,Arzoumanian2018a,Cromartie2019}. Upper limits are inferred through more elaborated interpretations of observational data indicating a finite remnant lifetime, e.g.~\cite{Lasky2014,Lawrence2015,Fryer2015,Margalit2017,Shibata2017,Rezzolla2018,Ruiz2018,Shunke2019,Margalit2019,Miller2020,Shao2020}. Estimates of the remnant's late-time behavior may be subject to considerable model dependencies. The present paper instead deals with the threshold for direct BH formation, which leads to different, relatively strong and clear observational features. Apart from implications for high-density matter physics~\cite{Oezel2016,Lattimer2016,Oertel2017},
$M_\mathrm{max}$ is also crucial for BH formation in core-collapse supernovae and BH physics~\cite{Sumiyoshi2006,Fischer2009,OConnor2011,Steiner2013,Fischer2014,Schneider2020}.

In this Letter we determine the impact of the properties of high-density matter on the threshold binary mass for prompt BH collapse. We go beyond current knowledge in several respects. We consider the so far largest sample of EoSs to determine $M_\mathrm{thres}$ and find new, tight relations describing its EoS dependence. For the first time, we systematically determine binary mass ratio effects on $M_\mathrm{thres}$. Furthermore, we investigate the impact of phase transitions on the collapse behavior.

We put forward four main applications of our new findings:

1) Along the lines of~\cite{Bauswein2013} we devise a more direct, more general and more accurate method to determine $M_\mathrm{max}$ from NS merger observations providing information about the
immediate merger outcome and about the total binary mass and combined tidal deformability from the inspiral GW signal.

2) We identify a new signature of a phase transition to deconfined quark matter in NS mergers. This stresses the enormous potential of future merger observations~\cite{Oechslin2004,Paschalidis2018,Most2019,Bauswein2019,Han2019,Christian2019,Sieniawska2018,Burgio2018a,Drago2018,Dexheimer2019,Chatziioannou2020,Chen2019,Alford2019,DePietri2019,Weih2019,Orsaria2019,Montana2019,Han2019a,Alvarez-Castillo2019,Li2020,Pereira2020,Blacker2020} to understand the phase diagram of matter in the non-perturbative regime of finite chemical potentials, which is not accessible by ab-initio QCD calculations~\cite{Bazavov2014,Borsanyi2014}. Currently, it is not known whether the hadron-quark phase transition takes place at typical NS densities. Identifying an imprint of the phase transition in merger observables will thus also provide invaluable insights for heavy-ion experiments, which will explore the phase diagram at such densities and finite temperature (but different isospin)~\cite{Friman2011,Blaschke2016,Hades2019}. 

3) Beyond these future prospects, our novel, more general relations are directly applicable in detection and analysis pipelines to quantify the likelihood of a specific merger outcome and thus for instance GW and kilonova characteristics~\cite{Abbott2017a,Dudi2018,Yang2018,Margalit2019,Coughlin2019a,Tsang2019,Metzger2019,Breschi2019,Foley2020,Agathos2020,Abbott2020,Chen2020,Krueger2020,Coughlin2020}. 

4) Furthermore, they are key input for current multimessenger constraints on NS properties as in~\cite{Bauswein2017,Radice2018,Radice2019,Koeppel2019,Bauswein2019a,Capano2020}.

See e.g.~\cite{Bauswein2017,Yang2018,Margalit2019,Tsang2019,Coughlin2019a,Koeppel2019,Breschi2019,Agathos2020,Foley2020,Capano2020,Paschalidis2019,Chen2020,Coughlin2020} for concrete implementations of $M_\mathrm{thres}$ dependencies. These applications can be significantly improved by the findings in this study. Below we describe our results mostly in the context of the first two applications.

{\it Simulations and setup:}
We perform three-dimensional relativistic hydrodynamical simulations of NS mergers for a large set of different EoSs of NS matter. For every EoS, we compute $M_\mathrm{thres}(\mathrm{EoS};q)$ for fixed binary mass ratios $q=M_1/M_2=1$ and $q=0.7$. Masses $M_1$ and $M_2$ of the individual binary components, $M_\mathrm{tot}$ and $M_\mathrm{thres}$ refer to the gravitational mass (for binaries at infinite orbital separation). Simulations start from quasi-equilibrium circular orbits a few revolutions before merger, with stars initially at zero temperature and in neutrino-less beta-equilibrium. The merger calculations are conducted with a relativistic smooth particle hydrodynamics code, which adopts the spatial conformal flatness condition to solve the Einstein field equations~\cite{Isenberg1980,Wilson1996}. More details on the simulation tool, comparisons to other codes (showing generally a very good agreement) and resolution studies can be found in~\cite{Oechslin2002,Oechslin2007,Bauswein2010,Bauswein2012a,Bauswein2013,Bauswein2014a,Bauswein2017,Koeppel2019,Agathos2020,Bauswein2020}.

In this study we consider 23 different hadronic EoSs~\cite{Banik2014,Fortin2018,Marques2017,Hempel2010,Typel2010,Typel2005,Alvarez-Castillo2016,Akmal1998,Goriely2010,Wiringa1988,Lattimer1991,Shen2011,Lalazissis1997a,Hempel2012,Douchin2001,Steiner2013,Muther1987,Alford2005,Engvik1996,Schneider2019,Read2009a}, which constitute our ``base sample'' and are consistent with astrophysical constraints from~\cite{Antoniadis2013,Abbott2017,Arzoumanian2018a}. To enlarge the parameter space, we optionally supplement those with 8 additional hadronic EoSs~\cite{Lackey2006,Glendenning1985,Shen2011,Lattimer1991,Lalazissis1997a,Sugahara1994a,Hempel2012,Toki1995} which are incompatible with the tidal deformability constraints from GW170817~\cite{Abbott2017}. Among all these EoSs five models include hyperons. Additonally, we consider 9 hybrid models with a first-order phase transition to deconfined quark matter leading to a strong softening of the EoS~\cite{Alvarez-Castillo2016,Kaltenborn2017,Bastian2018,Cierniak2018,Fischer2018,Bauswein2019}. These models vary in the onset density, the latent heat and the stiffness of quark matter~\cite{Bauswein2019,Blacker2020}. Among all 40 EoSs, 26  are fully temperature dependent. The remaining models are supplemented with an approximate treatment of thermal effects~\cite{Bauswein2010}. We refer to the Supplemental Material, which provides more details on the simulations and the different sets of EoS models and includes Refs.~\cite{Klaehn2015,Nambu1961,Klevansky1992,Klaehn2017,Balsara1995}. %only revised sentence for matching refs of suppl.%~5
We emphasize that our base sample covers well the full range of viable hadronic models.

\begin{figure}%[h]%fit1.py
\centering%[trim=left bottom right top, clip
\includegraphics[width=\columnwidth,trim=40 40 40 60,clip]{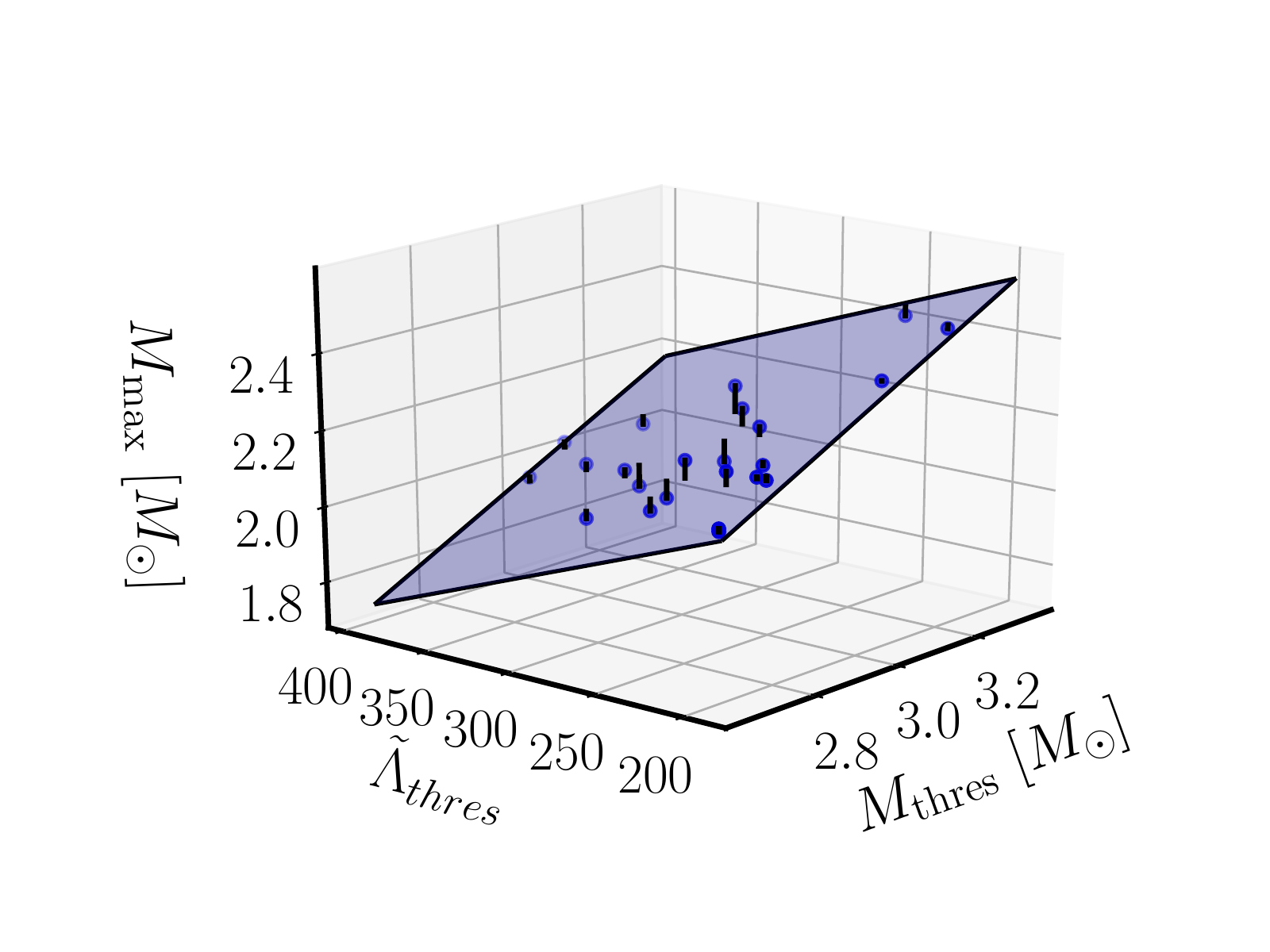}
\caption{Maximum mass $M_\mathrm{max}$ of nonrotating NSs as function of the threshold binary mass $M_\mathrm{thres}$ for prompt BH formation and combined tidal deformability $\tilde{\Lambda}_\mathrm{thres}=\tilde{\Lambda}(M_\mathrm{thres}/2)$ of the binary at the threshold for direct collapse. Blue plane shows a bilinear fit to the data for $q=1$. Short black lines visualize the deviation between fit (blue plane) and data points.}
\label{fig:1}
\end{figure}

{\it $M_\mathrm{max}$ determination:} 
We first consider results for the base sample of purely hadronic EoSs assuming that indications of a strong phase transition to quark matter may be independently provided by other observations or experiments~\cite{Friman2011,Fischer2018,Bauswein2019,Chen2019,Chatziioannou2020}. We further justify this assumption below by describing a new detectable signature of a phase transition.

Compiling the data for equal-mass mergers, Fig.~\ref{fig:1} reveals a tight relation between the maximum mass $M_\mathrm{max}$ of nonrotating NSs, the threshold binary mass $M_\mathrm{thres}$ and $\tilde{\Lambda}_\mathrm{thres}$. The latter is the combined tidal deformability of the binary system at the threshold mass, i.e. $\tilde{\Lambda}_\mathrm{thres}\equiv\Lambda(M_\mathrm{thres}/2)$ for $q=1$. $\tilde{\Lambda}$ describes the dominant EoS effects during the GW inspiral and is thus measurable~\cite{Read2009,DelPozzo2013,Read2013,Wade2014,Agathos2015,Chatziioannou2015,Hotokezaka2016,Abbott2017,Chatziioannou2018,De2018,Abbott2019}. It is defined by $\tilde{\Lambda}= \frac{16}{13(M_1+M_2)^5}((M_1+12 M_2)M_1^4\Lambda_1+(M_2+12 M_1)M_2^4\Lambda_2)$ with tidal deformabilities $\Lambda_{1(2)}$ of the individual binary components~\cite{Hinderer2008,Hinderer2010,Damour2010}. $\Lambda(M)$ is a stellar structure parameter and fully determined by the EoS through $\Lambda_{1(2)}=\frac{2}{3}k_2(R_{1(2)}/M_{1(2)})^5$ with the tidal Love number $k_2(M)$ and stellar radius $R(M)$ (factors of $G$ and $c$ suppressed).

The tidal deformability monotonically decreases with mass. Therefore, $\tilde{\Lambda}_\mathrm{thres}$ can be obtained from measurements of systems with different $M_\mathrm{tot}$ around $M_\mathrm{thres}$ through a simple interpolation. The tight relation in Fig.~\ref{fig:1} implies that a sufficiently accurate measurement of $M_\mathrm{thres}$ and $\tilde{\Lambda}_\mathrm{thres}$ determines the currently unknown maximum mass of nonrotating NSs. The data in Fig.~\ref{fig:1} is well described by a bilinear fit
\begin{equation}\label{eq:fit}
    M_\mathrm{max}(M_\mathrm{thres},\tilde{\Lambda}_\mathrm{thres})=a M_\mathrm{thres}+b\tilde{\Lambda}_\mathrm{thres}+c,
\end{equation}
with $a=0.632$, $b=-0.002~M_\odot$ and $c=0.802~M_\odot$. %fit1.py
The maximum residual of this fit is only 0.067~$M_\odot$, implying a potentially very accurate measurement of $M_\mathrm{max}$ (see Supplemental Material for fits with an enlarged set of EoSs). %(see Tab. \ref{tab:fit} for fits with an enlarged set of EoSs). 
The average deviation between Eq.~\eqref{eq:fit} and the underlying data is only 0.02~$M_\odot$. As an example, assuming $M_\mathrm{thres}$ to be measured within $0.05~M_\odot$ and $\tilde{\Lambda}_\mathrm{thres}$ within 5\%, an error propagation through Eq.~\eqref{eq:fit} yields $\Delta M_\mathrm{max}=0.06~M_\odot$.

For $q=0.7$ we obtain a similarly tight relation for hadronic EoSs with a maximum residual of 0.078~$M_\odot$ (with fit parameters $a=0.621$, $b=-0.001~M_\odot$, $c=0.582~M_\odot$; Tab.~II in Supplemental Material). For the same EoS $M_\mathrm{thres}$ of asymmetric systems is comparable to the one of equal-mass mergers (either equal or at most $0.2~M_\odot$ smaller). Moreover, we find the difference in $M_\mathrm{thres}$, i.e. $M_\mathrm{thres}(q=1)-M_\mathrm{thres}(q=0.7)$, to depend systematically on the EoS. % (fit 12 in Tab.~\ref{tab:fit}). 
See Supplemental Material for more details, a discussion of the systematic impact of the mass ratio and an intuitive explanation. %See the Supplemental Material for more details as well as an accompanying publication describing the systematic impact of the mass ratio in more detail and an intuitive explanation~\cite{Bauswein2020}.  -9

Based on our models we construct %provide 
additional bilinear fits (Tab.~II in Supplemental Material) quantifying their quality by the maximum residual and the average deviation between fit and data. For these relations we select different subsets of our data motivated by different assumptions on which additional information may be available (e.g. about $q$ or the presence of a phase transition). For instance, we consider only purely hadronic EoS models or a full set of EoSs including hybrid models with phase transitions, or we include binaries with a fixed mass ratio or a range in $q$. We also employ different independent variables, which may be measured more precisely in comparison to the quantities in Eq.~\eqref{eq:fit}. This includes (i) the chirp mass $\mathcal{M}_c=(M_1 M_2)^{3/5}/(M_1+M_2)^{1/5}$ if the mass ratio is not well constrained or strongly differs among the different events which are combined to determine $M_\mathrm{thres}$, (ii) the tidal deformability $\Lambda_{1.4}$ of a 1.4~$M_\odot$ NS, which may be more accurately and independently measured than $\tilde{\Lambda}_\mathrm{thres}$, or (iii) the radius $R_{1.6}$ of a 1.6~$M_\odot$ NS. We stress that cases (ii) and (iii), i.e. fits 8 to 11, are very promising when $\Lambda_{1.4}$ or $R_{1.6}$ are measured in a high SNR GW detection or by another astronomical observation, e.g. by NICER~\cite{Miller2019,Riley2019,Raaijmakers2019}.

Generally, all these choices lead to tight relations describing the collapse behavior. This is not unexpected considering the previously found relation $M_\mathrm{thres}=(-3.606\frac{G M_\mathrm{max}}{c^2 R_{1.6}}+2.38)M_\mathrm{max}$ for a smaller set of EoS models and only equal-mass mergers~\cite{Bauswein2013,Bauswein2017a}. NS radii are roughly constant in a considerable mass range around $M_\mathrm{thres}/2$ and the tidal deformability is known to scale approximately with NS radii (see also~\cite{Bauswein2017a} for a semi-analytic model of the collapse behavior). In comparison to previous results, the new relations presented here allow a more direct and more general implementation in analysis pipelines or waveform models because they involve quantities which are directly measurable from the GW inspiral (of the same event) and do not rely on additional information e.g. about $R_{1.6}$. They also include asymmetric binaries. We remark that the functional form of our new fits like Eq.~\eqref{eq:fit} is more physical compared to relations in~\cite{Bauswein2013}, which features a unphysical decrease of $M_\mathrm{thres}$ with $M_\mathrm{max}$ in a very small range of the parameter space. Finally, we directly compare the relations $M_\mathrm{max}(M_\mathrm{thres},\tilde{\Lambda}_\mathrm{thress})$ (Eq.~\eqref{eq:fit}) and $M_\mathrm{max}(M_\mathrm{thres},R_{1.6})$ (inverted relation from~\cite{Bauswein2013}). The maximum residual is 0.067~$M_\odot$ for the new relation compared to 0.26~$M_\odot$ for the latter. % (see also fit~10 in Tab.~\ref{tab:fit}). 
Hence, the relations describing the collapse behavior in this work are significantly more accurate while they even include more models and consider asymmetric mergers. 

Physically, relations as Eq.~\eqref{eq:fit} are understandable. $M_\mathrm{thres}$ is determined by two roughly independent EoS properties, namely $\tilde{\Lambda}_\mathrm{thres}$ characterizing the EoS stiffness at moderate densities and $M_\mathrm{max}$ at very high densities, both of which increase $M_\mathrm{thres}$. For fixed $M_\mathrm{thres}$ this implies that $\tilde{\Lambda}_\mathrm{thres}$ has to decrease with $M_\mathrm{max}$.

We emphasize that already a single measurement of $M_\mathrm{tot}$ and $\tilde{\Lambda}$ can yield a strong constraint on $M_\mathrm{max}$. Indications for a prompt collapse in a detection imply $M_\mathrm{tot}>M_\mathrm{thres}$ and $\tilde{\Lambda}<\tilde{\Lambda}_\mathrm{thres}$. From this follows through Eq.~\eqref{eq:fit} that the actual maximum mass of nonrotating NSs is smaller than $M_\mathrm{max}(M_\mathrm{tot},\tilde{\Lambda})$ (note the minus sign of the fit parameter $b$). If a measurement provides evidence for no direct BH formation, the maximum mass of NSs has to be larger than  $M_\mathrm{max}(M_\mathrm{tot},\tilde{\Lambda})$ because $M_\mathrm{tot}<M_\mathrm{thres}$ and $\tilde{\Lambda}>\tilde{\Lambda}_\mathrm{thres}$\footnote{For instance, a prompt (delayed) collapse event with $M_\mathrm{tot}=3.0~M_\odot$ and $150<\tilde{\Lambda}<250$ implies $M_\mathrm{max}<2.40~M_\odot$ ($M_\mathrm{max}>2.20$), which may be further tighten by incorporating additional $\tilde{\Lambda}$ data from other events.}.

{\it Further applications:} All aforementioned relations %in Tab.~\ref{tab:fit} 
are bilinear and thus easy to invert for other applications requiring for instance $M_\mathrm{thres}$ or the tidal deformability to be the dependent quantity (applications 3 and 4). We stress that one can exploit our different relations describing the collapse behavior even if some parameters are poorly constrained as for instance in~\cite{Bauswein2017,Koeppel2019,Capano2020} yielding a lower bound on NS radii of about 11~km. 

Our models also show that the range of $\tilde{\Lambda}_\mathrm{thres}$ is relatively large: for equal-mass mergers $200\lesssim\tilde{\Lambda}_\mathrm{thres}\lesssim450$, whereas $200\lesssim\tilde{\Lambda}_\mathrm{thres}\lesssim650$ for $q=0.7$~\cite{Bauswein2020}, which is significantly broader than previously assumed (cf.~\cite{Zappa2018,Agathos2020,Bernuzzi2020}). Hence, only for $\tilde{\Lambda}<200$ a prompt collapse can be assumed, while depending on $q$ only events with $\tilde{\Lambda}\gtrsim650$ may safely be classified as no direct collapse. This is for example relevant for kilonova observations and GW searches to determine whether there may be contributions from strong postmerger GW emission. These ranges imply that independent of $M_\mathrm{max}$ the tidal deformability of a 1.37~$M_\odot$ NS has to be larger than about 200 following the arguments in Ref.~\cite{Margalit2017,Bauswein2017,Radice2018,Bauswein2019a} favoring a delayed collapse in GW170817. This limit is less than the one reported in~\cite{Radice2018,Radice2019}, but our data clearly shows that current observations do not exclude EoSs with $\Lambda_{1.37}>200$ in line with~\cite{Bauswein2017,Kiuchi2019,Bauswein2019a}.

\begin{figure}
\centering
\includegraphics[width=\columnwidth]{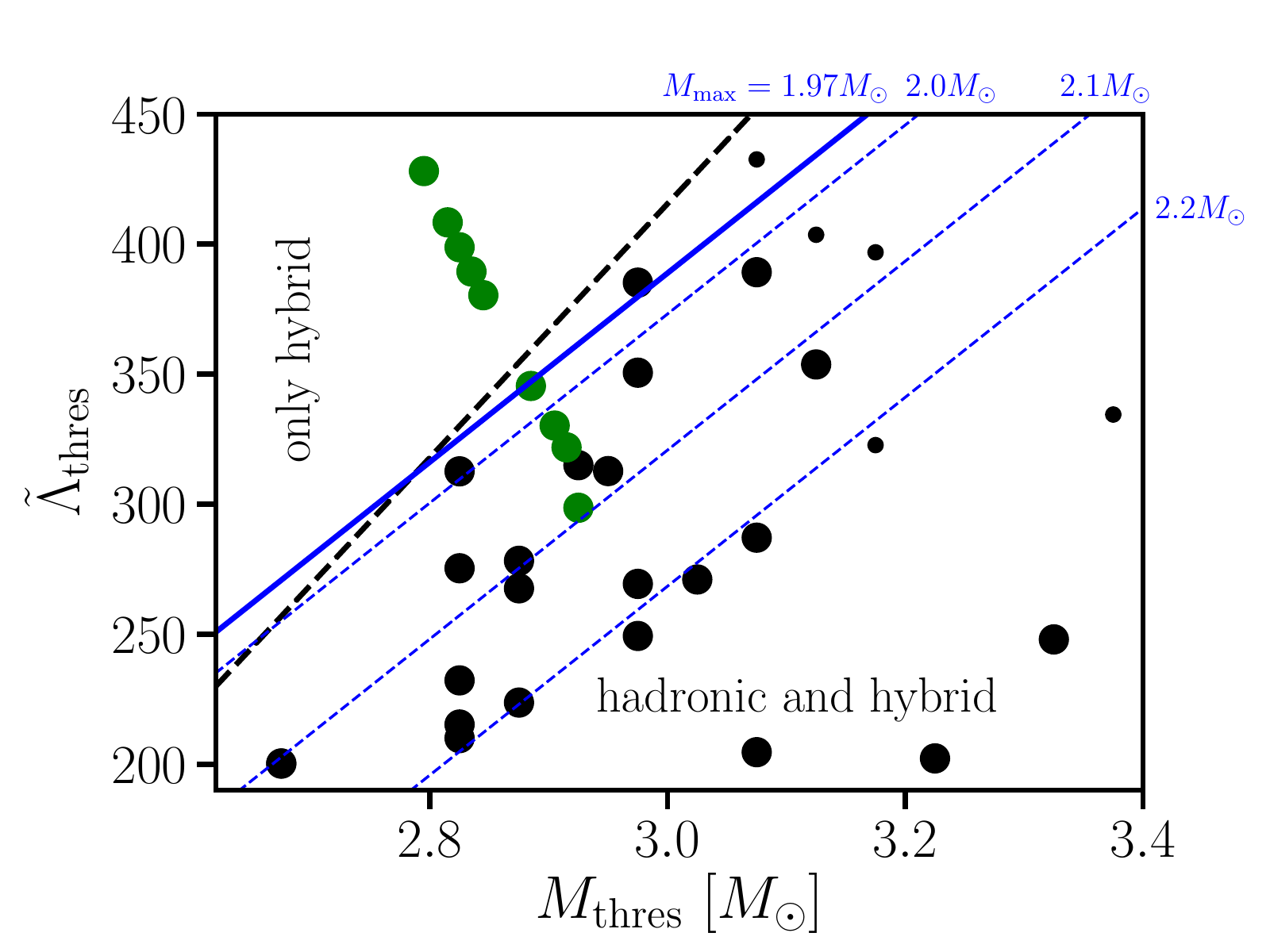}
\caption{Combined tidal deformability of binaries at the threshold to prompt BH formation as function of threshold binary mass $M_\mathrm{thres}$ for direct collapse for different hybrid EoSs (green points) and purely hadronic EoSs (black points; small symbols are models of the ``excluded'' sample, three outside the plot range). Dashed line indicates boundary beyond which only hybrid models exist (Eq.~\eqref{eq:hybrid}). In all except for one hybrid model the phase transition occurs after merger. Hence, $\tilde{\Lambda}_\mathrm{thres}=\Lambda(M_\mathrm{thres}/2)$ is that of purely hadronic stars, which for our models all are described by the same hadronic EoS below the onset density of the phase transition. Therefore, $\tilde{\Lambda}_\mathrm{thres}(M_\mathrm{thres})$ of these hybrid models appear to line up on a single curve, following $\Lambda(M)$ of the hadronic EoS. The blue lines show curves of constant $M_\mathrm{max}$ using fit~\eqref{eq:fit} for purely hadronic EoSs explaining the absence of viable hadronic models in the upper left corner.}
\label{fig:2}
\end{figure}

%Combined tidal deformability of binaries at the threshold to prompt BH formation as function of the threshold binary mass $M_\mathrm{thres}$ for direct collapse for different hybrid EoSs (green points) and purely hadronic EoSs (black points; small symbols are models for which the EoS is incompatible with the tidal deformability inference from GW170817, three of such EoSs at $M_\mathrm{thres}>3.4~M_\odot$ are not shown). Overplotted crosses mark hyperonic EoSs, whereas the red plus sign displays the ALF2 EoS, where quark matter resembles properties of hadronic matter~\cite{Alford2005,Read2009a}. Dashed line indicates a boundary beyond which only hybrid models exist (Eq.~\eqref{eq:hybrid}). In all except for one hybrid model the phase transition occurs after merger. Hence, $\tilde{\Lambda}_\mathrm{thres}=\Lambda(M_\mathrm{thres}/2)$ is that of purely hadronic stars, which for our models all are described by the same hadronic EoS below the onset density of the phase transition. Therefore, $\tilde{\Lambda}_\mathrm{thres}(M_\mathrm{thres})$ of these hybrid models appear to line up on a single curve, following $\Lambda(M)$ of the hadronic EoS. The blue lines show curves of constant $M_\mathrm{max}$ using fit~\eqref{eq:fit} for purely hadronic EoSs. This explains the absence of viable hadronic models in the upper left corner.

{\it New signature of phase transition:}
By additionally considering the results with hybrid EoSs, we identify a new observable signature of the hadron-quark phase transition, which may occur in NSs. Figure~\ref{fig:2} shows $\tilde{\Lambda}_\mathrm{thres}$ as function of $M_\mathrm{thres}$ for all EoSs with $q=1$. It is striking that all 31 purely hadronic models are located below the dashed line given by
\begin{equation}\label{eq:hybrid}
  \tilde{\Lambda}_\mathrm{thres}^\mathrm{hybrid}= 488 (M_\mathrm{thres}/M_\odot)-1050,
\end{equation}
whereas most hybrid models with a phase transition occur above this curve, i.e. at relatively small $M_\mathrm{thres}$ but larger $\tilde{\Lambda}_\mathrm{thres}$. Hence, a combined measurement of $(M_\mathrm{thres},\tilde{\Lambda}_\mathrm{thres})$ with $\tilde{\Lambda}_\mathrm{thres}> 488 (M_\mathrm{thres}/M_\odot)-1050$ provides strong evidence for the presence of a phase transition.

A strong phase transition induces a softening of the EoS at higher densities and thus destabilizes the merger product, i.e. yields a relatively small $M_\mathrm{thres}$. For most of these models $M_\mathrm{thres}/2$ is smaller than the smallest mass $M_\mathrm{onset}$ at which quark matter appears in nonrotating NSs. Hence, the inspiralling stars are purely hadronic and the corresponding tidal deformability $\tilde{\Lambda}_\mathrm{thres}$ does not carry any information about the phase transition and is thus relatively large. To some extent this effect is comparable to results in~\cite{Bauswein2019}, where a stronger compactification of the merger remnant by the phase transition leads to a characteristic increase of the postmerger GW frequency.

Figure~\ref{fig:2} is a projection of the data point of Fig.~\ref{fig:1} onto the $M_\mathrm{thres}-\tilde{\Lambda}_\mathrm{thres}$ plane. We can thus draw lines of constant $M_\mathrm{max}$ using fit~\eqref{eq:fit} for purely hadronic EoSs. This explains why no viable purely hadronic EoS models occur above the dash line: Only models with $M_\mathrm{max}<1.97~M_\odot$ could yield a $(M_\mathrm{thres},\tilde{\Lambda}_\mathrm{thres})$ combination in the upper left corner, which however is excluded by pulsar mass measurements~\cite{Antoniadis2013,Cromartie2019}. The situation is different for EoSs with a strong phase transition even if they yield a maximum mass above~2~$M_\odot$. As explained those models can feature a strong softening at higher densities, which leads to a destabiliation of the merger remnant and a correspondingly low $M_\mathrm{thres}$. A low $M_\mathrm{thres}$ implies that the merging stars of the system with $M_\mathrm{tot}=M_\mathrm{thres}$ are hadronic and relatively light. This leads to a relatively large $\tilde{\Lambda}_\mathrm{thres}$. Thus, hybrid models with this behavior can lead to strong deviations from the $M_\mathrm{max}(M_\mathrm{thres},\tilde{\Lambda}_\mathrm{thres})$ relation of hadronic EoSs and the data points $(M_\mathrm{thres},\tilde{\Lambda}_\mathrm{thres})$ can occur in a regime inaccessible by viable hadronic models.

This explains our finding and solidifies that the described signature through the criterion in Eq.~\eqref{eq:hybrid} is indicative of a phase transition. For $q=0.7$ we find a qualitatively similar behavior~\cite{Bauswein2020}.

Following this argumentation we further point out that the limit which indicates a phase transition can be updated when pulsar measurements increase the lower bound on $M_\mathrm{max}$.

We stress several advantages of this new signature to uncover the hadron-quark phase transition in NS mergers. (1) $\tilde{\Lambda}_\mathrm{thres}$ does not need to be determined with very high precision in comparison to the accuracy which would be required to detect a relatively weak kink at $M_\mathrm{onset}$ in $\Lambda(M)$ indicating a phase transition (e.g. Fig.~3 in~\cite{Bauswein2019a}). A precision of 10\% to 30\% is sufficient. (2) For most hybrid EoSs studied here $M_\mathrm{thres}/2<M_\mathrm{onset}$, which implies that $\tilde{\Lambda}_\mathrm{thres}$ is larger than $\Lambda(M_\mathrm{onset})$ and thus easier to measure (because of stronger finite-size effects and possibly more frequent systems). Detecting a phase transition with high $M_\mathrm{onset}$ becomes increasingly challenging for methods employing only the GW inspiral~\cite{Chatziioannou2020,Chen2019}, in which case our signature is particularly promising because it is sensitive to the very high-density regime. (3) Already a single measurement with a constraint on $(M_\mathrm{thres},\tilde{\Lambda}_\mathrm{thres})$ may reveal indications of a phase transition. (4) $M_\mathrm{tot}$ can be measured with very good precision and there are a number of different signals potentially revealing the merger product, e.g. postmerger GWs, kilonovae, and possibly gamma-ray bursts, implying that a sufficient $M_\mathrm{thres}$ determination is conceivable in the near future. In fact, all in principle required observables have already been measured~\cite{Abbott2017,Abbott2017b}.

Notably, not all hybrid EoSs lie in the ``hybrid regime'' above the dashed line in Fig.~\ref{fig:2}. These are models with a very strong stiffening of the EoS in the quark phase (with $M_\mathrm{max}$ exceeding the one of the purely hadronic reference model; see Supplemental Material). One may refer to this as a coarse variant of the masquerade problem~\cite{Alford2005}, where hybrid models roughly resemble the mass-radius relation of purely hadronic EoSs. The stiffening leads to a stabilization of the merger product and thus to a relatively large $M_\mathrm{thres}$ and consequently a relatively small $\tilde{\Lambda}_\mathrm{thres}\equiv\Lambda(M_\mathrm{thres}/2)$. A $(M_\mathrm{thres},\tilde{\Lambda}_\mathrm{thres})$ below the dashed curve does thus generally not allow to infer the nature of high-density NS matter. %However, the hadronic models slightly below the dashed line are those which feature a transition to hyperonic matter. Hence, 
However, the proximity to the dash curve indicates a softening of the EoS at higher densities and possibly the occurrence of a weak phase transition.

{\it Conclusions:} Future work should investigate potential systematic uncertainties which might exist on a very low level, overcome those by improved numerical and physical modeling, and explore in more detail the observational features resulting from the collapse behavior. Also, an even larger set of hybrid EoSs should be considered because our current models vary the properties of the quark phase but employ the same hadronic EoS at densities below the phase transition. This hadronic reference model lies in the middle of the range given by current astrophysical and experimental constraints~\cite{Danielewicz2002,Tsang2018,Lattimer2013,Oertel2017,Krueger2013,Antoniadis2013,Arzoumanian2018a,Abbott2017,Bauswein2017,De2018,Abbott2018}. We thus expect that other hybrid models show the same behavior; such models should essentially be shifted parallel to the dashed line.% An even larger set of EoSs may be used to refine the boundary between the pure ``hybrid regime'' and the mixed ``hadronic and hybrid regime''. 

\acknowledgements{Acknowledgements: We thank G. Martinez-Pinedo and H.-T. Janka for helpful discussions. We thank C. Constantinou, M. Prakash, A. Schneider and J. Smith for help with their EoS tables.  AB acknowledges support by the European Research Council (ERC) under the European Union's Horizon 2020 research and innovation programme under grant agreement No. 759253. A.B. and S.B. acknowlege support by Deutsche Forschungsgemeinschaft (DFG, German Research Foundation) - Project-ID 279384907 - SFB 1245. A.B. and V.V. acknowledge support by DFG - Project-ID 138713538 - SFB 881 (``The Milky Way System'', subproject A10). D.B. and T.F. acknowledge support from the Polish National Science Center (NCN) under grant no. 2019/33/B/ST9/03059. T.F. acknowledges support from NCN under Grant No. UMO-2016/23/B/ST2/00720. NUFB acknowledges support from NCN under grant number 2019/32/C/ST2/00556. N.S. is supported
by the ARIS facility of GRNET in Athens (SIMGRAV, SIMDIFF\ and BNSMERGE allocations)
and the ``Aristoteles Cluster'' at AUTh, as well as by the COST actions CA16214
``PHAROS'', CA16104 ``GWVerse'', CA17137 ``G2Net'' and CA18108 ``QG-MM''.}

%\bibliography{references}

\input{main.bbl}
\end{bibunit}

% Supp for arxiv

\clearpage
\newpage
\begin{bibunit}%[supp]
\setcounter{figure}{0}
\setcounter{equation}{0}
\setcounter{page}{1}

\onecolumngrid
\begin{center}
{\large \bf Supplemental Material}
\vspace*{1cm}
\twocolumngrid
\end{center}

\section{Equation of state sample, simulations and fit formulae} \label{sec:eos}

In this letter, we describe the collapse behavior, i.e. the threshold mass for prompt BH formation, for a large sample of EoS models. We perform three-dimensional relativistic hydrodynamical simulations as in~\cite{Oechslin2002,Oechslin2007,Bauswein2010,Bauswein2012a,Bauswein2013} to determine $M_\mathrm{thres}$. All these simulations make the following assumptions about the initial data. The NSs have an irrotational velocity field, i.e. no intrinsic spin, and the stellar matter is at zero temperature. The composition (electron fraction) is given by neutrino-less beta-equilibrium. We start calculations a few cycles before merging and assume a quasi-circular orbit. We explicitly refer to these earlier publications for more details and additional information.

By considering calculations with different $M_\mathrm{tot}$ and determining the respective merger product, we obtain $M_\mathrm{thres}$ with an accuracy of at least $\pm 0.025~M_\odot$. For every EoS we vary $M_\mathrm{tot}$ in steps of $0.05~M_\odot$ and define the threshold mass as $M_\mathrm{thres}=0.5 (M_\mathrm{tot,delayed}+M_\mathrm{tot,prompt})$. Within our set of simulations  $M_\mathrm{tot,delayed}$ ($M_\mathrm{tot,prompt}$) is the binary mass of the most (least) massive system leading to a delayed (prompt) collapse. (We define a prompt-collapse event as those systems where the minimum lapse function $\alpha_\mathrm{min}$ continuously decreases and never increases after merging. This is a meaningful definition because an increasing $\alpha_\mathrm{min}$ implies a bounce of the merging binary components, which leads to an increase of the ejecta mass and thus to a fundamentally different electromagnetic signal compared to a prompt collapse.) % {\bf raus?} We also note that the lifetime of the remnant increases rapidly with decreasing total mass. Therefore, the range of $M_\mathrm{tot}$ leading to events with very short remnant lifetime, and thus weaker postmerger GW emission, is very narrow. Hence, a potential misclassification of a merger with a delayed collapse after a very short lifetime as prompt-collapse event does not significantly bias the determination of $M_\mathrm{thres}$.)  
We follow this prescription to determine $M_\mathrm{thres}$ for fixed mass ratios of $q=M_1/M_2=0.7$ and $q=1$.

In total we consider 40 EoS models (see Tab.~\ref{tab:eos}), which we group in three subsets:

(a) The ``base sample'' consists of 23 purely hadronic EoSs which are compatible with current astrophysical constraints from pulsar mass measurements and from limits on the tidal deformability in GW170817. We require the maximum mass of nonrotating NSs to be larger than 1.97~$M_\odot$, which is the lower bound of the error bars from~\cite{Antoniadis2013,Cromartie2019}, and we require the tidal deformability of a 1.37~$M_\odot$ NS to be smaller than 800, which is the less stringent limit from an analysis of finite-size effects during the inspiral of GW170817~\cite{Abbott2017}. Generally, we prefer to be less restrictive with regard to possible constraints to include as many models as possible for a sufficient coverage of the parameter space.

(b) An extended hadronic sample includes additional 8 hadronic models which are incompatible with the aforementioned measurements (at the two sigma and 90\% confidence level, respectively). We refer to this set as ``excluded hadronic sample'' which is useful to cover even more models and understand dependencies.

(c) The ``hybrid sample'' comprises a set of 9 different EoSs which feature a phase transition to deconfined quark matter beyond some transition density~\cite{Fischer2018}. This sample serves to investigate the impact of a phase transition on the threshold binary mass for prompt collapse. Most of these models have been employed in~\cite{Bauswein2019}, where additional information can be found\footnote{One of the models within our hybrid sample has not been described previously. The vector-interaction enhanced bag model (vBAG) has been derived from the Schwinger-Dyson formalism of QCD as limiting case for a particular choice of the gluon propagator (for details see Ref.~\cite{Klaehn2015}). VBAG features chiral symmetry restoration via a chiral bag constant, in which aspect vBag resembles thermodynamic results of commonly used EoSs of the Nambu-Jona-Lasinio type~\cite{Nambu1961,Klevansky1992}. Furthermore, (de)confinement is taken into account through an additional bag constant which is directly linked to properties of the underlying hadronic EOS \cite{Hempel2010,Typel2010,Hempel2012} at the chiral symmetry restoration. This approach ensures the simultaneous restoration of chiral symmetry and (de)confinement~\cite{Klaehn2017}. Also, one model of the DD2F-SF family of hybrid EoSs (DD2F-SF8) was not included in~\cite{Bauswein2019}, which is why we here provide its specific parameters, namely $\sqrt{D_0}=240$~MeV, $\alpha=0.1~\mathrm{fm^6}$, $a=0.0~\mathrm{MeV\,fm^3}$, $b=0.0~\mathrm{MeV\,fm^9}$, $c=0.0~\mathrm{fm^6}$, $\rho_1=80~\mathrm{MeV\,fm^3}$ (see~\cite{Bastian2018,Bauswein2019}).}. These models are based on a single hadronic EoS below the transition density, but differ in the properties of the quark matter phase. This leads to different onset densities of the phase transition, different latent heat (density jump across the phase transition) and  different stiffness of the quark phase EoS. All hybrid models are fully temperature dependent. This is important because also the phase boundaries vary with temperature, which cannot be easily captured by a simplified treatment of thermal effects (see e.g.~\cite{Bauswein2010}).

\input{eostable.tex}

Note that sample (a) includes the ALF2 EoS implemented as piecewise polytrope~\cite{Alford2005,Read2009a}. This EoS is formally a hybrid model with a transition to quark matter. However, it is build such that it resembles the properties of hadronic matter, which is why we count it for the hadronic sample. The base sample contains three EoSs with a transition to hyperonic matter, and the excluded hadronic sample includes another two of such models.

All EoS models are listed in Tab.~\ref{tab:eos}, which includes the references for each EoS. We indicate in Tab.~\ref{tab:eos} to which subset a given EoS model belongs. The table also includes different stellar parameters, which we employ to characterize the EoSs: the maximum mass $M_\mathrm{max}$ of a nonrotating NS, the radius of a nonrotating NS with 1.6~$M_\odot$ and the tidal deformablity of a 1.4~$M_\odot$ NS.

26 EoSs of our total sample are implemented in the form of tables, which include the temperature and composition dependence. For the remaining models, which provide the EoS only at $T=0$, we use an approximate prescription of the thermal pressure, setting the tunable thermal ideal-gas index $\Gamma_\mathrm{th}=1.75$ (see~\cite{Bauswein2010}). The temperature dependent models are marked with a ``T'' in Tab.~\ref{tab:eos}, barotropic EoSs are indicated by ``B''.

\begin{figure}% RM-mthres.py
\centering
\includegraphics[width=\columnwidth]{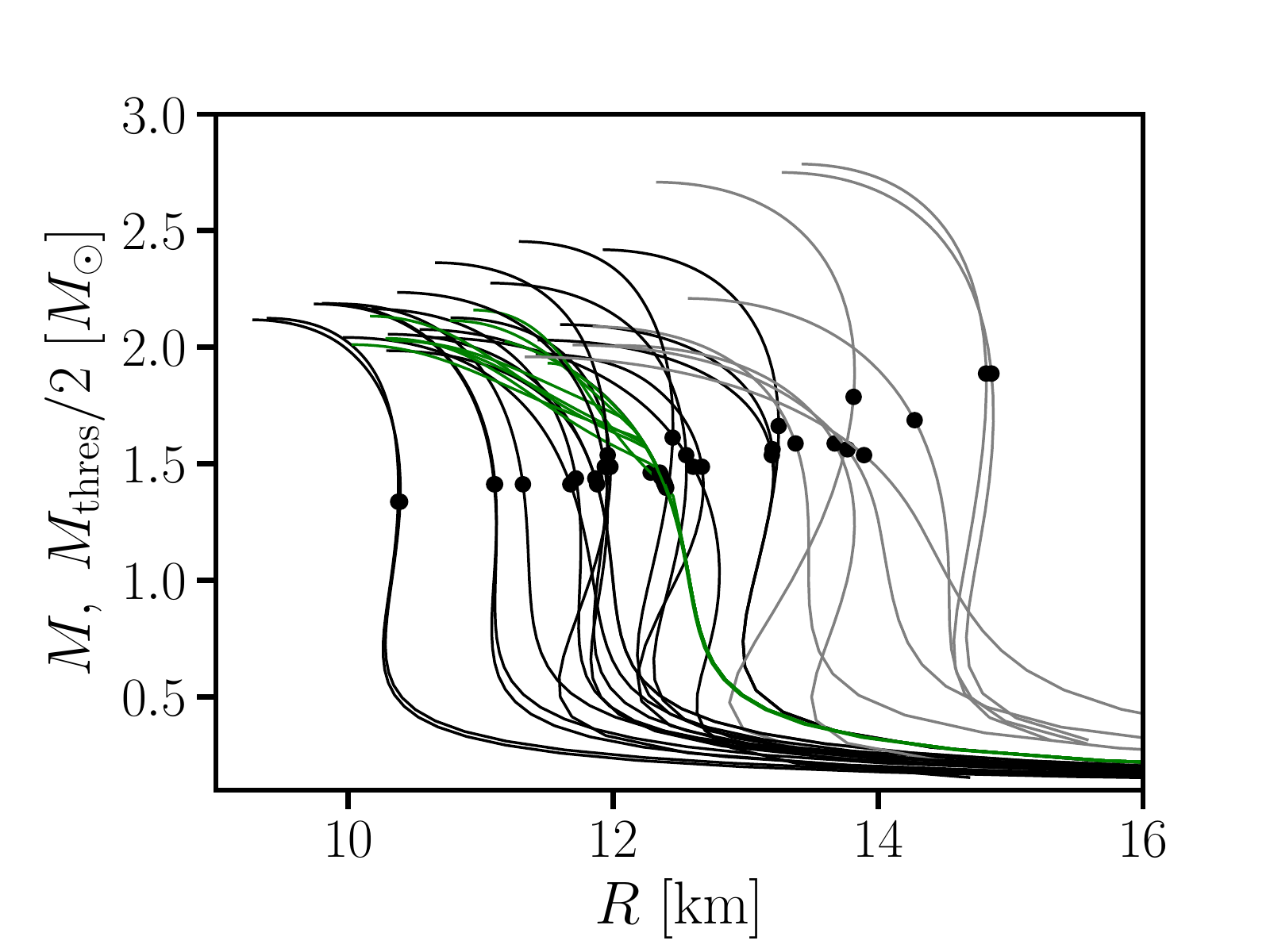}
\caption{Mass-radius relations for the EoS models employed in this study. Black lines display the purely hadronic base sample, while gray curves show purely hadronic EoSs which are incompatible with current astrophysical constraints. Models of the hybrid sample are plotted in green. The dots visualize  $M_\mathrm{thres}/2$ of the respective EoS. See main text for more explanations.} 
\label{fig:MR}
\end{figure}

Figure~\ref{fig:MR} shows the resulting mass-radius relations for all EoSs employed in this study. The hadronic base sample is displayed by black lines and the excluded hadronic sample by gray lines. The green lines depict the hybrid models featuring a characteristic kink at $M_\mathrm{onset}$, which is the smallest mass where quark matter is present (see also Fig.~2 in~\cite{Bauswein2019} for a zoom-in). Figure~\ref{fig:MR} demonstrates that our set of models covers the full viable range of hadronic EoSs with regard to the range of stellar parameters. This is crucial because in the main text we argue that hadronic models are constrained to a certain area in the $M_\mathrm{thres}-\tilde{\Lambda}_\mathrm{thres}$ plane, dubbed ``mixed regime''. It is thus reasonable to expect that any other hadronic EoS will also follow this behavior. It should approximately resemble one of the EoSs from our sample which is the most similar one to it. Note that there is a unique relation between the M-R relations and the EoS for zero-temperature models. Hence, also the viable range of thermodynamical properties, explicitly $P(\rho)$ is well sampled by our set of EoS models.

As mentioned, the hybrid models in our sample employ a single hadronic model at lower densities. In this context we recall that this hadronic EoS is fully compatible with current astrophysical and nuclear physics constraints (see~\cite{Danielewicz2002,Tsang2018,Lattimer2013,Oertel2017,Krueger2013,Antoniadis2013,Arzoumanian2018a,Abbott2017,Bauswein2017,De2018,Abbott2018}). In fact, it roughly falls in the middle of the ranges given by these measurements. We thus expect that variations to this model do not lead to a fundamentally different behavior.

The dots in Fig.~\ref{fig:MR} visualize the threshold mass for prompt BH formation for the different EoSs with a binary mass ratio $q=1$. We plot $M_\mathrm{thres}/2$ on the mass-radius relation of the corresponding EoS at $(R(M_\mathrm{thres}/2),M_\mathrm{thres}/2)$. The dots thus show the radii $R_\mathrm{thres}=R(M_\mathrm{thres}/2)$ of the inspiraling stars before merging. The figure illustrates that the radii $R_{1.6}$ of 1.6~$M_\odot$ NSs are well suited to characterize the relevant EoS regime of binaries close to the threshold for prompt collapse. Note that densities increase during merging. Hence, the collapse behavior is also affected by the EoS at higher densities than those which are realized in stars with $M_\mathrm{thres}/2$. This explains that the combination $R_{1.6}$ and $M_\mathrm{max}$ represents a good choice to characterize the collapse behavior.% One also recognizes that $R_{1.6}$ is roughly constant around  $M_\mathrm{thres}/2$, which means that 

Table~\ref{tab:eos} provides the threshold masses for equal-mass binaries and for asymmetric binaries with a mass ratio $q=0.7$. In addition, we list $\tilde{\Lambda}_\mathrm{thres}$ as the combined tidal deformability of the binary with the threshold mass for every EoS model and mass ratio considered in this study. Note that for equal-mass binaries $\tilde{\Lambda}_\mathrm{thres}=\Lambda(M_\mathrm{thres}/2)$.

\input{fittablewapr.tex}

Based on the data in Tab.~\ref{tab:eos} we construct different bilinear fit formulae describing the collapse behavior of NS mergers, which are discussed in the main paper. These relations connect $M_\mathrm{max}$, $M_\mathrm{thres}$ and one more stellar parameter characterizing the EoS. We provide these relations in Tab.~\ref{tab:fit} with the maximum mass $M_\mathrm{max}$ being the dependent variable. We emphasize that these fits are bilinear. It is thus trivial to obtain relations with $M_\mathrm{thres}$ (or any other quantity) being the dependent variable, which may be useful for many applications (see main paper). The various choices for the fit functions and the underlying data set are motivated by different assumptions on which quantities may be measured or constrained. The quality of the relations is quantified by the maximum residual and the average deviation between fit and data.

% Generally, all fits are found to be relativly accurate, and the tightness depends on 
% 
% These relations include the threshold binary mass $M_\mathrm{thres}$ for prompt BH formation 
% 
% --

Note that $M_\mathrm{thres}$ for some of the EoSs which have already been considered in~\cite{Bauswein2013} slightly differ from the values reported therein. The reasons are that here we determine $M_\mathrm{thres}$ with higher accuracy, i.e. finer sampling in $M_\mathrm{tot}$, and we slightly modified the treatment of artificial viscosity within the smooth particle hydrodynamics scheme by implementing an additional factor for lowering viscosity in a pure shear flow~\cite{Balsara1995}.

We finally remark that the results presented in Tabs.~\ref{tab:eos} and~\ref{tab:fit} constitute the largest study of the collapse behavior of binary mergers to date. It includes the largest set of EoS models and it determines for the first time systematically the threshold mass for asymmetric mergers. In this study we intentionally do not include additional data from other groups, which are publicly available~\cite{Hotokezaka2011,Zappa2018,Koeppel2019,Agathos2020}. First, this would not enlarge our sample significantly. Second, for this study it is important to work with a consistent set of data to quantify the quality of fit relations. Other studies determine $M_\mathrm{thres}$ with different accuracy and, moreover, it is difficult to assess intrinsic model dependencies of other simulation results like for instance the resolution dependence, EoS implementation or residual orbital eccentricity. Generally, there is a very good agreement comparing our results to the ones from other groups~\cite{Hotokezaka2011,Zappa2018,Koeppel2019,Agathos2020}. This said we stress that more future work will be required to fully understand the impact of the numerical treatment and different physical effects on $M_\mathrm{thres}$. Clearly, our new findings highlight the scientific value of such future efforts.

\begin{figure*}%suppMthres-q-DD2F.py; suppMthres-q-sfhx.py
\centering
\includegraphics[width=\columnwidth]{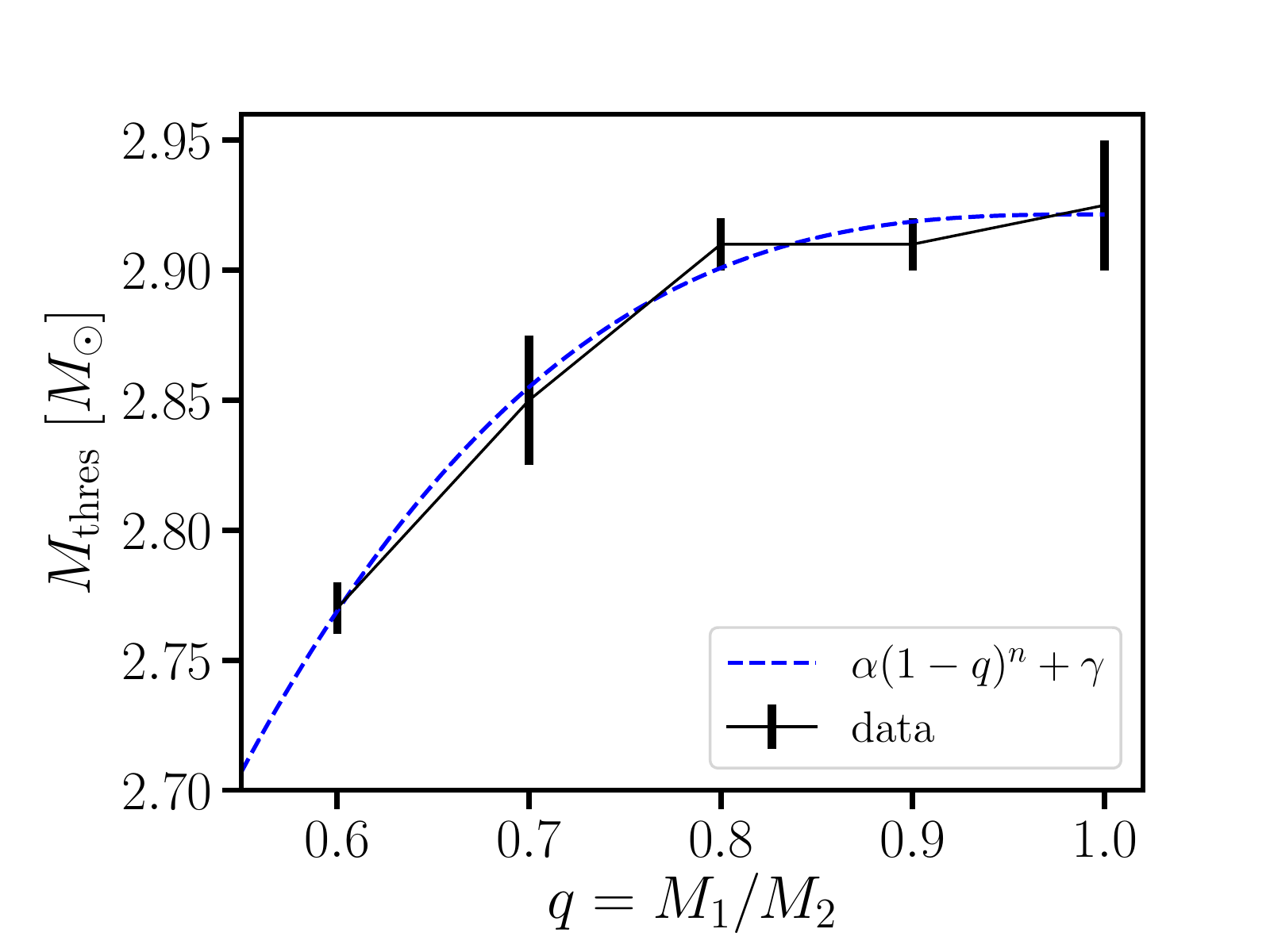}
\includegraphics[width=\columnwidth]{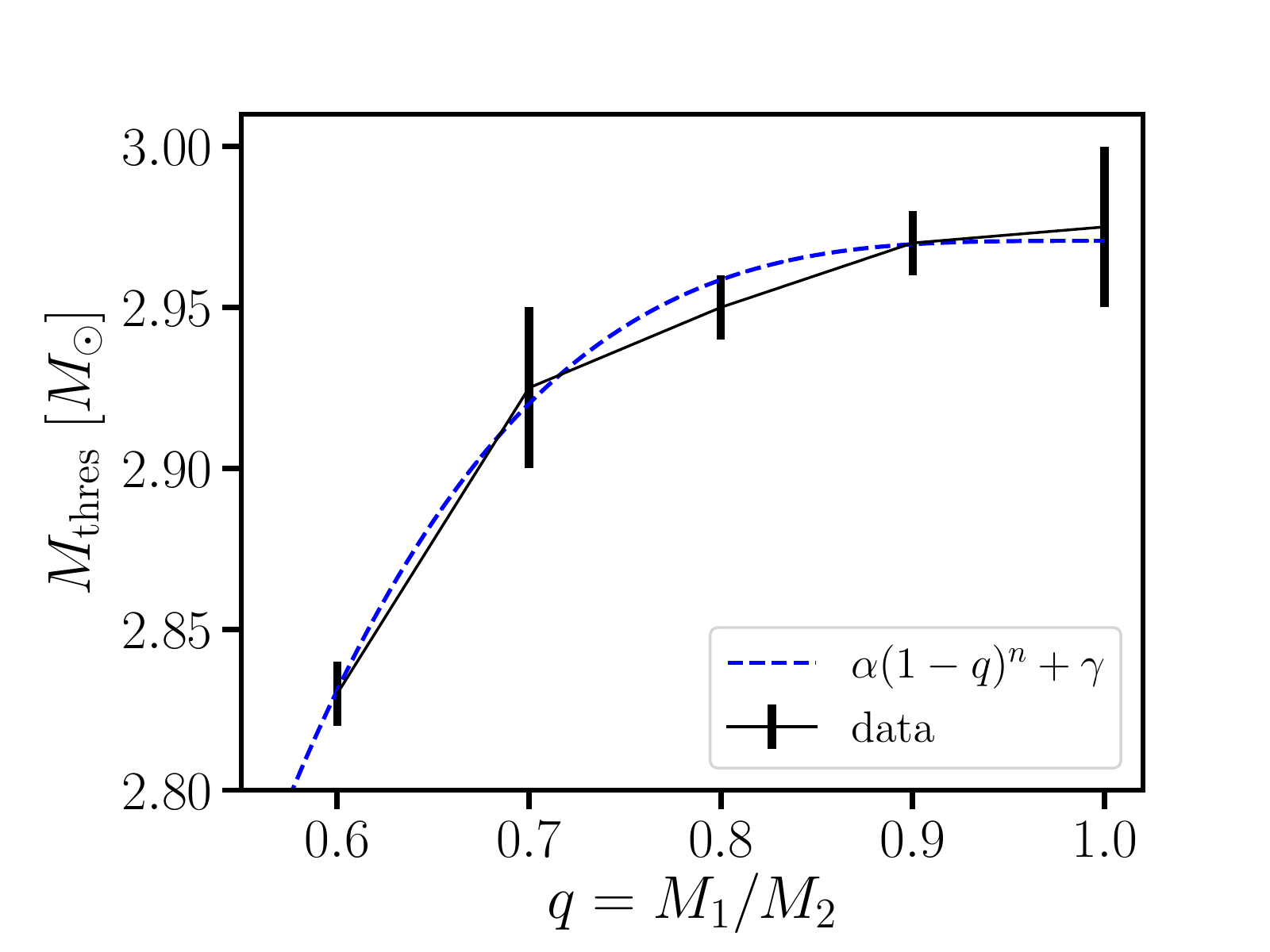}
\caption{Threshold binary mass for prompt BH formation as function of the binary mass ratio $q$ for the DD2F EoS (left panel) and the SFHX EoS (right panel). The width of the error bar indicates the accuracy to which $M_\mathrm{thres}$ has been determined for the given $q$ (see main text). The dashed blue curve shows a least-squares fit of the form $M_\mathrm{thres}(q)=\alpha(1-q)^n+\gamma$.}
\label{fig:Mthrq}
\end{figure*}

\section{Impact of binary mass ratio}
In this study we determine the threshold binary mass for fixed mass ratios of $q=0.7$ and $q=1$, which is the range inferred for GW170817. Since the sensitivity of current GW instruments continues to increase, future merger observations will reveal the binary mass with higher accuracy at the same distance, whereas the mass ratio of events at larger distance will not be obtained with good precision. Therefore, we construct fit formulae for fixed mass ratios as well as for a range of mass ratios.

Determining $M_\mathrm{thres}$ for a range in $q$ is important for observations where the mass ratio is not known very well. Obviously, in simulations the threshold mass can only be computed for fixed mass ratios. Hence, the range of $M_\mathrm{thres}$ for $0.7\leq q\leq 1$ is determined by individual models. We explicitly assume that $M_\mathrm{thres}$ varies monotonically with $q$, such that  $M_\mathrm{thres}(q=1)$ and $M_\mathrm{thres}(q=0.7)$ are sufficient to specify the range. While this is a very reasonable assumption and physically intuitive, we confirm this by additional calculations for selected EoS models. 

Figure~\ref{fig:Mthrq} shows $M_\mathrm{thres}(q)$ for the DD2F and SFHX EoSs, where we explicitly calculate  $M_\mathrm{thres}$ for $q=\{0.6,0.7,0.8,0.9,1.0\}$. The error bars specify the precision to which  $M_\mathrm{thres}$ was determined in the simulations (with the upper edge being $M_\mathrm{tot,prompt}$ and the lower edge being $M_\mathrm{tot,delayed}$ as described above). Within the given accuracy the calculations confirm that $M_\mathrm{thres}(q)$ is indeed a monotonic function of the mass ratio $q$. Note that the dependence on $q$ is not precisely linear but follows approximately a higher-order polynomial. By a fit assuming a dependence $(1-q)^n$ we determine a power of $n=2.89$ for DD2F and $n=3.53$ for SFHX. The impact of the mass ratio thus becomes stronger for stronger binary mass asymmetry. For small deviations from $q=1$ the threshold mass is roughly constant. We find a qualitatively similar behavior in additional simualtions for the DD2 and SAPR EoSs. Our observations are in line with previous calculations for $q=0.9$ in~\cite{Bauswein2013} and for $q=0.6$ in~\cite{Bauswein2017}. These conclusions are also consistent with the simulations for fixed mass in~\cite{Bernuzzi2020}, which however do not directly determine $M_\mathrm{thres}$.

Figure~\ref{fig:Mthrq} and the data in Tab.~\ref{tab:eos} show a very clear dependence on the binary mass ratio, namely, generally, a decrease of $M_\mathrm{thres}$ with asymmetry. This general behavior is physically understandable based on Newtonian point particles. For the same total mass and the same orbital distance, circular orbits of asymmetric binaries have less angular momentum than equal-mass systems. Hence, the available angular momentum to support the merger remnant is reduced for $q<1$ leading to smaller $M_\mathrm{thres}$ (assuming the merging to take place at the same orbital distance).

\begin{figure}% Dmthres-r16-mmax-computed.py
\centering
\includegraphics[width=\columnwidth,trim=40 40 40 60,clip]{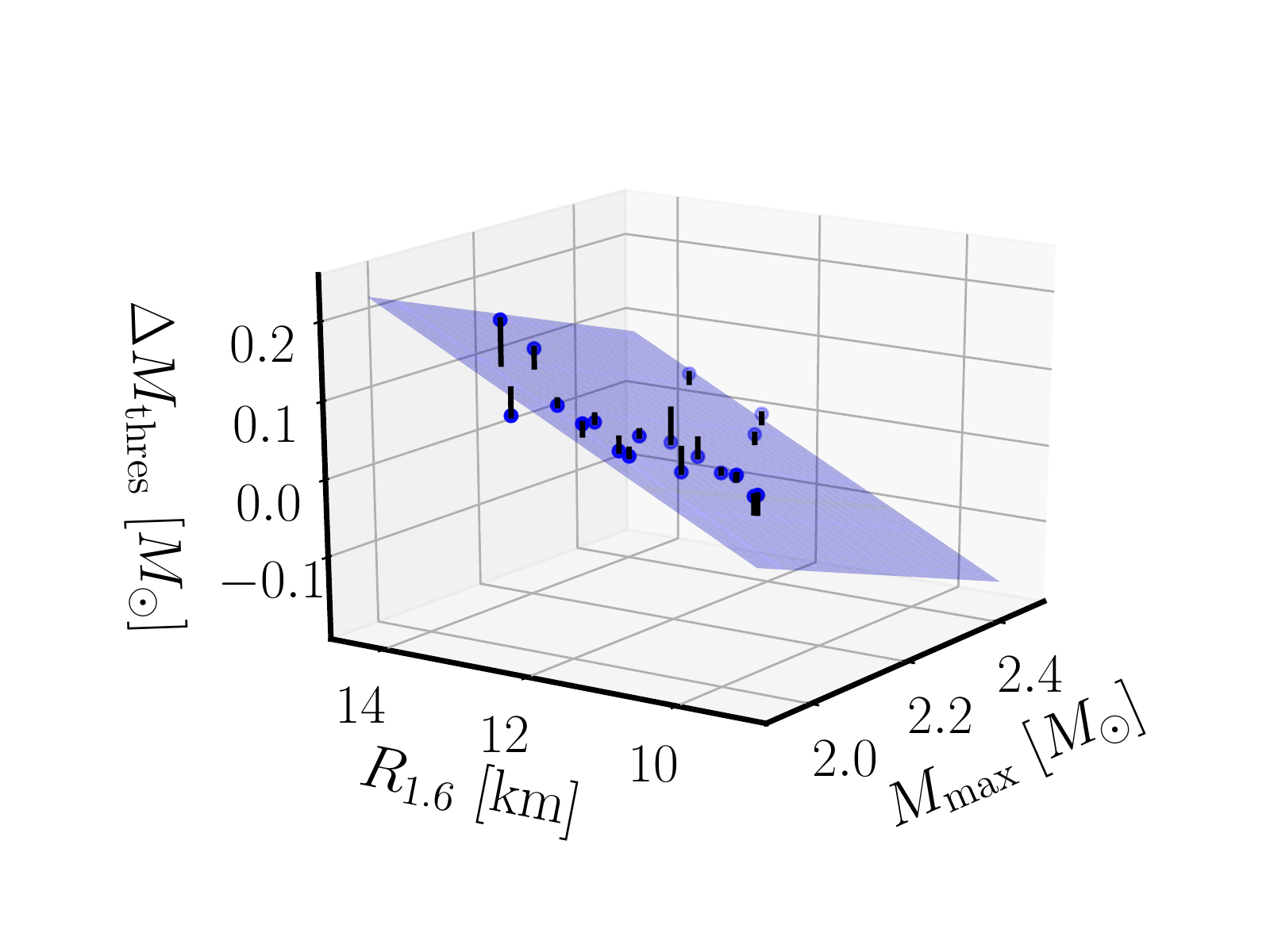}
\caption{Impact of the binary mass ratio on the collapse behavior of NS mergers for the hadronic base sample. Blue dots display $\Delta M_\mathrm{thres}$ as the difference between the threshold mass for prompt collapse of equal-mass binaries and of asymmetric binaries with $q=0.7$ as function of the maximum mass $M_\mathrm{max}$ and radius $R_{1.6}$ of 1.6~$M_\odot$ NSs. The blue plane is a bilinear fit to the data. Short black lines visualize the deviations between fit and underlying data.}
\label{fig:DMthr}
\end{figure}

We finally comment on a finding that we already highlight in the main paper. The impact of the mass ratio on  $M_\mathrm{thres}$ is differently strong for different EoSs (see last fit in Tab.~\ref{tab:fit}). Importantly, also the difference between  $M_\mathrm{thres}(q=1)$  and  $M_\mathrm{thres}(q=0.7)$ follows a specific EoS dependence, which can be well described by $\Delta M_\mathrm{thres} = M_\mathrm{thres}(q=1) - M_\mathrm{thres}(q=0.7)= a M_\mathrm{max}+ b R_{1.6}+ c$ with the fit parameters given in Tab.~\ref{tab:fit}. Figure~\ref{fig:DMthr} shows this relation for $\Delta M_\mathrm{thres}$ and demonstrates its tightness.

A more extended discussion of mass ratio effects on the collapse behavior will be presented in a forthcoming publication. Here we still note that this particular dependence of $\Delta M_\mathrm{thres}$ also explains the findings of~\cite{Kiuchi2019}, which tentatively indicate that for soft EoSs which yield very small NS radii, the remnants of asymmetric mergers may be more stable than that of equal-mass binaries of the same total mass\footnote{Note that~\cite{Kiuchi2019} runs only simulations for a fixed total binary mass without determining $M_\mathrm{thres}$.}. A stronger stabilization of remnants resulting from asymmetric mergers would seem somewhat in tension with the results from~\cite{Bauswein2013,Bauswein2017,Bernuzzi2020}, which show a destabilization for stiffer EoS models if $q$ deviates from unity. All these different findings however become fully consistent in the light of the particular EoS dependence revealed by Fig.~\ref{fig:DMthr}. It clearly shows that for very soft EoSs $M_\mathrm{thres}(q=1)$  and  $M_\mathrm{thres}(q=0.7)$ are very comparable and that the threshold mass for asymmetric mergers may even be larger than the one of equal-mass binaries in a small parameter range.

\input{supp.bbl}
%\bibliography{references}

\end{bibunit}

\end{document}

%% file: journalheader.tex
\newcommand{\apjl}{Astrophys. J. Lett.}%
\newcommand{\apjs}{Astrophys. J. Supp.}%
\newcommand{\aap}{Astron. Astrophys.}%
\newcommand{\mnras}{Mon. Not. Roy. Astron. Soc.}%
\newcommand{\lrr}{Living Reviews in Relativity}%
\newcommand{\pasa}{Publications of the Astronomical Society of Australia}
\newcommand{\prx}{Phys. Rev. X}
\newcommand{\araa}{Annual Review of Astronomy and Astrophysics}
\newcommand{\physrep}{Physics Reports}

%% file: eostable.tex
\begin{table*} 
\begin{tabular}{|l|l|c|c|c|c|c|c|c|c|c|c|}\hline

EoS & T/B & $M_\mathrm{max}$ & $R_{1.6}$ & $\Lambda_{1.4}$ & 
$M_\mathrm{thres}{(q=1)}$ & 
$\tilde{\Lambda}_\mathrm{thres}{(q=1)}$ &
$M_\mathrm{thres}{(q=0.7)}$ &
$\tilde{\Lambda}_\mathrm{thres}{(q=0.7)}$ &
sample &
Ref. \\

& & $(M_\odot)$ & $(\mathrm{km})$ & & $(M_\odot)$ & & $(M_\odot)$ & & & \\ \hline

BHBLP  &  T  & 2.098 & 13.192 & 691.0 & 3.125 & 353.8 & 2.975 & 512.8 &  b & \cite{Banik2014}  \\ \hline
DD2Y  &  T  & 2.031 & 13.169 & 691.0 & 3.075 & 389.2 & 2.875 & 622.1 &  b & \cite{Fortin2018,Marques2017}  \\ \hline
DD2  &  T  & 2.419 & 13.247 & 694.8 & 3.325 & 248.0 & 3.275 & 300.3 &  b & \cite{Hempel2010,Typel2010}  \\ \hline
DD2F  &  T  & 2.077 & 12.220 & 423.1 & 2.925 & 315.0 & 2.850 & 427.7 &  b & \cite{Typel2005,Typel2010,Alvarez-Castillo2016}  \\ \hline
APR  &  B  & 2.187 & 11.253 & 245.9 & 2.825 & 232.2 & 2.825 & 260.2 &  b & \cite{Akmal1998}  \\ \hline
BSK20  &  B  & 2.165 & 11.648 & 317.4 & 2.875 & 267.6 & 2.875 & 300.3 &  b & \cite{Goriely2010}  \\ \hline
eosUU  &  B  & 2.189 & 11.057 & 227.9 & 2.825 & 215.2 & 2.825 & 241.1 &  b & \cite{Wiringa1988}  \\ \hline
LS220  &  T  & 2.041 & 12.478 & 537.0 & 2.975 & 350.6 & 2.875 & 519.0 &  b & \cite{Lattimer1991}  \\ \hline
LS375  &  T  & 2.709 & 13.767 & 950.8 & 3.575 & 223.5 & 3.575 & 248.5 &  e & \cite{Lattimer1991}  \\ \hline
GS2  &  T  & 2.089 & 13.369 & 717.2 & 3.175 & 322.7 & 3.025 & 487.3 &  e & \cite{Shen2011}  \\ \hline
NL3  &  T  & 2.787 & 14.795 & 1360.3 & 3.775 & 228.5 & 3.775 & 257.9 &  e & \cite{Hempel2010,Lalazissis1997a}  \\ \hline
Sly4  &  B  & 2.043 & 11.523 & 292.4 & 2.825 & 275.4 & 2.775 & 352.8 &  b & \cite{Douchin2001}  \\ \hline
SFHO  &  T  & 2.056 & 11.751 & 331.5 & 2.875 & 278.2 & 2.825 & 352.9 &  b & \cite{Steiner2013}  \\ \hline
SFHOY  &  T  & 1.986 & 11.748 & 331.5 & 2.825 & 312.6 & 2.725 & 441.5 &  b & \cite{Fortin2018,Marques2017}  \\ \hline
SFHX  &  T  & 2.127 & 11.963 & 393.1 & 2.975 & 269.3 & 2.925 & 328.3 &  b & \cite{Steiner2013}  \\ \hline
TM1  &  T  & 2.210 & 14.347 & 1142.0 & 3.375 & 334.5 & 3.225 & 525.0 &  e & \cite{Sugahara1994a,Hempel2012}  \\ \hline
TMA  &  T  & 2.008 & 13.660 & 928.0 & 3.175 & 396.9 & 2.975 & 698.1 &  e & \cite{Toki1995,Hempel2012}  \\ \hline
BSK21  &  B  & 2.276 & 12.543 & 511.4 & 3.075 & 287.1 & 3.075 & 317.7 &  b & \cite{Goriely2010}  \\ \hline
GS1  &  T  & 2.750 & 14.864 & 1392.1 & 3.775 & 229.6 & 3.775 & 260.4 &  e & \cite{Shen2011}  \\ \hline
eosAU  &  B  & 2.125 & 10.357 & 149.9 & 2.675 & 200.3 & 2.675 & 222.2 &  b & \cite{Wiringa1988}  \\ \hline
WFF1  &  B  & 2.118 & 10.362 & 150.0 & 2.675 & 200.2 & 2.675 & 220.1 &  b & \cite{Wiringa1988,Read2009a}  \\ \hline
WFF2  &  B  & 2.186 & 11.048 & 222.4 & 2.825 & 210.0 & 2.825 & 235.3 &  b & \cite{Wiringa1988,Read2009a}  \\ \hline
MPA1  &  B  & 2.454 & 12.448 & 475.9 & 3.225 & 202.2 & 3.225 & 224.6 &  b & \cite{Muther1987,Read2009a}  \\ \hline
ALF2  &  B  & 1.973 & 12.616 & 565.1 & 2.975 & 385.2 & 2.875 & 510.1 &  b & \cite{Alford2005,Read2009a}  \\ \hline
H4  &  B  & 2.010 & 13.716 & 846.4 & 3.125 & 403.6 & 2.925 & 699.6 &  e & \cite{Lackey2006,Read2009a}  \\ \hline
DD2F-SF-1  &  T  & 2.134 & 12.141 & 423.1 & 2.845 & 380.4 & 2.770 & 497.8 &  h & \cite{Kaltenborn2017,Bastian2018,Fischer2018,Bauswein2019}  \\ \hline
DD2F-SF-2  &  T  & 2.160 & 12.061 & 421.2 & 2.925 & 298.6 & 2.870 & 399.3 &  h & \cite{Kaltenborn2017,Bastian2018,Fischer2018,Bauswein2019}  \\ \hline
DD2F-SF-3  &  T  & 2.032 & 12.189 & 423.1 & 2.825 & 398.8 & 2.720 & 570.1 &  h & \cite{Kaltenborn2017,Bastian2018,Fischer2018,Bauswein2019}  \\ \hline
DD2F-SF-4  &  T  & 2.029 & 12.220 & 423.1 & 2.835 & 389.5 & 2.725 & 566.9 &  h & \cite{Kaltenborn2017,Bastian2018,Fischer2018,Bauswein2019}  \\ \hline
DD2F-SF-5  &  T  & 2.038 & 11.928 & 423.1 & 2.815 & 408.4 & 2.725 & 539.2 &  h & \cite{Kaltenborn2017,Bastian2018,Fischer2018,Bauswein2019}  \\ \hline
DD2F-SF-6  &  T  & 2.012 & 12.219 & 423.1 & 2.795 & 428.1 & 2.675 & 635.5 &  h & \cite{Kaltenborn2017,Bastian2018,Fischer2018,Bauswein2019}  \\ \hline
DD2F-SF-7  &  T  & 2.115 & 12.220 & 423.1 & 2.905 & 330.2 & 2.825 & 451.2 &  h & \cite{Kaltenborn2017,Bastian2018,Fischer2018,Bauswein2019}  \\ \hline
DD2F-SF-8  &  T  & 2.025 & 12.216 & 422.3 & 2.915 & 321.9 & 2.810 & 467.3 &  h & \cite{Kaltenborn2017,Bastian2018,Fischer2018,Bauswein2019}  \\ \hline
VBAG  &  T  & 1.932 & 12.214 & 422.3 & 2.885 & 345.5 & 2.775 & 505.4 &  h & \cite{Cierniak2018}  \\ \hline
ENG  &  B  & 2.236 & 11.899 & 367.5 & 2.975 & 249.3 & 2.975 & 279.7 &  b & \cite{Engvik1996,Read2009a}  \\ \hline
APR3  &  B  & 2.363 & 11.954 & 364.8 & 3.075 & 204.6 & 3.075 & 228.1 &  b & \cite{Akmal1998,Read2009a}  \\ \hline
GNH3  &  B  & 1.959 & 13.756 & 850.4 & 3.075 & 432.6 & 2.875 & 799.3 &  e & \cite{Glendenning1985,Read2009a}  \\ \hline
SAPR  &  T  & 2.194 & 11.462 & 265.7 & 2.875 & 223.7 & 2.875 & 254.5 &  b & \cite{Schneider2019}  \\ \hline
SAPRLDP  &  T  & 2.247 & 12.369 & 449.3 & 3.025 & 271.0 & 3.025 & 309.4 &  b & \cite{Schneider2019}  \\ \hline
SSkAPR  &  T  & 2.028 & 12.304 & 442.6 & 2.950 & 312.7 & 2.875 & 420.8 &  b & \cite{Schneider2019}  \\ \hline

\end{tabular}
\caption{EoS employed in this study. Second column indicates whether EoS table provides temperature dependence (T) or whether table is barotropic and supplemented by an approximate temperature treatment (B). Next three columns list stellar parameters which characterize the EoS, i.e. maximum mass $M_\mathrm{max}$, radius $R_{1.6}$ of a 1.6~$M_\odot$ NS and tidal deformability of a 1.4~$M_\odot$ NS. Next four columns provide threshold binary mass $M_\mathrm{thres}$ for prompt collapse and combined tidal deformability at $M_\mathrm{thres}$ for equal-mass mergers and asymmetric mergers with a binary mass ratio $q=0.7$. Penultimate entry classifies to which of the three EoS samples the given model belongs, where ``b'' stands for ``base sample'', ``e'' for ``excluded hadronic sample'' and ``h'' for ``hybrid sample''. Last column gives reference of EoS model.}
\label{tab:eos}
\end{table*}

%% file: fittablewapr.tex
\begin{table*}
\begin{tabular}{|l|l|c|c|c|c|c|c|c|}\hline
no. & fit & EoS sample & $q$ & $a$ & $b$ & $c$ & max. dev. & av. dev. \\ \hline
1 & $M_\mathrm{max}=a M_\mathrm{thres}+b\tilde{\Lambda}_\mathrm{thres} +c$ & base sample & 1.0    &
0.632 & -1.866e-03 & 0.802 & 0.067 & 0.023  \\ \hline   %fit1.py

1e& $M_\mathrm{max}=a M_\mathrm{thres}+b\tilde{\Lambda}_\mathrm{thres} +c$ & base sample + 8 excl. had.\footnote{We include 8 hadronic EoS incompatible with~\cite{Abbott2017}. } & 1.0             & 
0.63 & -2.002e-03 & 0.841 & 0.106 & 0.031  \\ \hline   %fit1we.py

2 & $M_\mathrm{max}=a M_\mathrm{thres}+b\tilde{\Lambda}_\mathrm{thres} +c$ & base sample & 0.7       &
0.621 & -6.637e-04 & 0.582 & 0.078 & 0.023  \\ \hline  %fit2.py

3 & $M_\mathrm{max}=a M_\mathrm{thres}+b\tilde{\Lambda}_\mathrm{thres} +c$ & base sample & 1.0 and 0.7 &
0.53 & -7.409e-04 & 0.833 & 0.153 & 0.051   \\ \hline  %fit3.py

4 & $M_\mathrm{max}=a M_\mathrm{thres}+b\tilde{\Lambda}_\mathrm{thres} +c$ & base sample + 9 hyb. & 1.0 &
0.477 & -1.156e-03 & 1.077 & 0.138 & 0.054  \\ \hline  %fit4.py

5 & $M_\mathrm{max}=a M_\mathrm{thres}+b\tilde{\Lambda}_\mathrm{thres} +c$ & base sample + 4 hyb.\footnote{We include hybrid models with  $(M_\mathrm{thres},\tilde{\Lambda}_\mathrm{thres})$ below the dashed line in Fig.~2 in the main paper.} & 1.0          & 
0.627 & -1.840e-03 & 0.811 & 0.089 & 0.028  \\ \hline  %fit5.py

6 & $M_\mathrm{max}=a \mathcal{M}_\mathrm{c,thres}+b\tilde{\Lambda}_\mathrm{thres} +c$ & base sample & 1.0 and 0.7    & 
1.073 & -6.956e-04 & 1.018 & 0.166 & 0.057  \\ \hline  %fit6.py

7 & $M_\mathrm{max}=a \mathcal{M}_\mathrm{c,thres}+b\tilde{\Lambda}_\mathrm{thres} +c$ & base sample + 9 hyp & 1.0 and 0.7 & 
0.899 & -4.680e-04 & 1.167 & 0.203 & 0.066  \\ \hline   % fit7.py

8 & $M_\mathrm{max}=a M_\mathrm{thres}+b\Lambda_{1.4} +c$ & base sample & 1.0  & 
1.47 & -1.166e-03 & -1.714 & 0.08 & 0.039 \\ \hline  % fit8.py

9 & $M_\mathrm{max}=a M_\mathrm{thres}+b\Lambda_{1.4} +c$ & base sample & 0.7  & 
1.052 & -5.709e-04 & -0.671 & 0.072 & 0.03 \\ \hline  % fit10.py

10& $M_\mathrm{max}=a M_\mathrm{thres}+b R_{1.6} +c$ & base sample & 1.0 & 
1.685 & -2.761e-01 & 0.488 & 0.078 & 0.029  \\ \hline  % fit9.py

11& $M_\mathrm{max}=a M_\mathrm{thres}+b R_{1.6} +c$ & base sample & 0.7 & 
1.143 & -1.318e-01 & 0.412 & 0.07 & 0.021  \\ \hline  % fit11.py

%ne, alt ist alt. 9 & old fit??? & had. & 1.0                                              & 1.655 & -2.659452e-01 & 0.451 & 0.077 & 0.029  \\ \hline
12 & $M_\mathrm{thres}^{q=1}-M_\mathrm{thres}^{q=0.7}=a M_\mathrm{max}+b R_{1.6} +c$ & base sample & 1.0 and 0.7 &
-0.285 & 4.859e-02 & 0.079 & 0.061 & 0.019 \\ \hline %Dmthres-r16-mmax-computed.py
   
%13 & $\Lambda_{1.8}=a M_\mathrm{thres} + b$ & base sample + 9 hyb. & 1.0 & 
%192 & -498 &          & 25 & 7.4 \\ \hline %lam18-mthres-computed.py

%14 & $R_{1.8}=a M_\mathrm{thres} + b$ & base sample + 9 hyb. & 1.0  & 
%4.332 & -0.917 & & 0.615 & 0.252 \\ \hline %r18-mthres-computed.py

\end{tabular}
\caption{Different bilinear fits describing the collapse behavior (see main text). Third and fourth columns list the data set employed for the fit specifying the sample of EoSs and the binary mass ratio $q$. $a$, $b$ and $c$ are fit parameters. Last two columns provide the maximum and average deviation between fit and the underlying data. All units are such that masses are in $M_\odot$ and radii in~km; $\Lambda$ is dimensionless.}% Last two rows list noteworthy bivariate linear fits.}%$\Lambda_{1.4}$ is the tidal deformability of a NS with 1.4~$M_\odot$, $R_{1.6}$ is the radius of a NS with 1.6~$M_\odot$.
\label{tab:fit}
\end{table*}